\documentclass[12pt]{article}

\usepackage{graphics,cite,amssymb,epsfig,float,psfrag}
\usepackage[usenames,dvips]{color}
\usepackage{rotating}

\newcommand{\lsim}{\raisebox{-0.13cm}{~\shortstack{$<$ \\[-0.07cm] $\sim$}}~} 
\newcommand{\gsim}{\raisebox{-0.13cm}{~\shortstack{$>$ \\[-0.07cm] $\sim$}}~} 
 
\newcommand{\tb}{\tan\beta} 
\newcommand{\beq}{\begin{eqnarray}} 
\newcommand{\eeq}{\end{eqnarray}} 
\newcommand{\s}{\\ \vspace*{-4mm}}
\begin{document}

\vspace{1cm}

\hfill LPT--ORSAY--11/27

\hfill CERN--PH--TH/2011--069

\vspace*{1.5cm}

\begin{center}

{\large\bf Implications of the ATLAS and CMS searches in the channel}

\vspace{.1cm}

{\large\bf $\mathbf{ pp \! \to \! Higgs\! \to\! \tau^+\tau^-}$ for the MSSM
and SM Higgs bosons}

\vspace*{.8cm}

{\large Julien Baglio$^{1}$ and Abdelhak Djouadi$^{1,2}$} 

\vspace*{8mm}

$^1$ Laboratoire de Physique Th\'eorique, Universit\'e Paris XI et CNRS,
F-91405 Orsay, France.\\
$^2$ Theory Unit, CERN, 1211  Gen\`eve 23, Switzerland.
\end{center}

\vspace{1.4cm}

\begin{abstract} 

We discuss the implications of the recent constraints on the Higgs sector of the
Minimal Supersymmetric extension of the Standard Model obtained by the ATLAS and
CMS collaborations at the lHC with $\sqrt s=7$ TeV and 36 pb$^{-1}$ of data. The
main production and detection channel that is relevant in these analyses is the
gluon--gluon and bottom quark fusion mechanisms leading to neutral Higgs bosons
which subsequently decay into tau lepton pairs, $gg, b\bar b\! \to\! {\rm
Higgs}\! \to\! \tau^+\tau^-$. In this note, we show that: $i)$ the exclusion
limits are in fact more general than indicated by the ATLAS and CMS analyses and
are essentially independent of the scenario for the supersymmetric particle
spectrum; $ii)$ when the exclusion limits are applied to the lowest theory
prediction for the Higgs production cross section times branching ratio, when
all theoretical uncertainties are  taken into account, the bounds are somewhat
less stringent; $iii)$ the exclusion limits from the $pp \to$Higgs$\to \tau^+
\tau^-$ process are so strong that only a modest improvement would be possible
when other MSSM Higgs detection channels are considered, even with femtobarn
level accumulated data. Finally and most important, we point out that the
prospects for the search for the Standard Model Higgs boson in the inclusive
$gg\to H\to \tau^+\tau^-$ channel, that is not currently considered  by the
ATLAS and CMS collaborations, turn out to be very promising and with a few
inverse femtobarn data it might provide a convincing discovery signal in the
difficult 115--135 GeV mass range for the standard Higgs boson. 

\end{abstract} 

\newpage

\subsection*{1. Introduction} 

The first analyses of supersymmetric Higgs production at the early stage of the
large Hadron Collider (lHC) have been recently released by the ATLAS  and CMS
collaborations \cite{ATLAS,CMS}. Searches for the neutral Higgs bosons of the 
Minimal Supersymmetric  Standard Model (MSSM), in which the Higgs sector is
extended to contain five scalar particles, two CP--even $h,H$ bosons, a CP--odd
or pseudoscalar $A$ boson and two charged $H^\pm$ particles \cite{Review}, have
been performed at a center of mass energy of $\sqrt  s=7$ TeV and with 36
pb$^{-1}$ of data in  the inclusive channel  $pp \!\to\!  \tau^+\tau^-$. The
obtained results  are rather impressive: in the absence of an additional  Higgs
contribution  on top of the continuum  background, very stringent limits on the
MSSM Higgs sector, beyond those available from the LEP \cite{LEP} and Tevatron
\cite{Tevatron} experiments, have been derived. In particular, values $\tb \gsim
25$--40 for  the ratio of vacuum expectation values of the two Higgs fields
(that is expected to lie in the range $1 \lsim \tb \lsim 60$) have been excluded
for  pseudoscalar Higgs  mass values in the  range between 100 and 200 GeV. \s

In the MSSM, only two parameters are needed to describe the Higgs sector at
tree--level:  $\tb$ and the pseudoscalar mass $M_A$. At high $\tb$ values, $\tb
\gsim 10$, one of the neutral CP--even states has almost exactly the properties
of the Standard Model (SM)  Higgs particle: its couplings to fermions and gauge
bosons are the same, but its mass is restricted to values $M_h^{\rm max} 
\approx 110$--135 GeV depending on the radiative corrections  that enter the
MSSM Higgs sector  \cite{Review,RCHiggs}. The other CP--even state ($H$ in the
decoupling regime $M_A \gsim M_h^{\rm max}$ and $h$ in the antidecoupling regime
$M_A \lsim M_h^{\rm max}$) and the CP--odd state, that we will denote
collectively by $\Phi\!=\!A,H(h)$, are then almost degenerate in mass and have
the same very strongly enhanced couplings to bottom quarks and $\tau$--leptons
($\propto \tb$) and suppressed couplings to  top quarks and gauge bosons. This
leads to a rather simple phenomenology for these states: the $\Phi$ bosons decay
almost exclusively into $b\bar b$ and $\tau^+\tau^-$ pairs  and,
at hadron colliders, these states are primarily produced  in the gluon--gluon
fusion mechanism, $gg \to \Phi$, which dominantly  proceeds through $b$--quark
triangular loops \cite{ggH-LO,SDGZ}  and  bottom--quark fusion, $b\bar b \to
\Phi$  \cite{bbH-NLO,bbH-NNLO}, in which the bottom quarks are directly taken
from the protons in a five active flavor scheme\footnote{This  process is
similar to the $p p \to b\bar b\Phi$ channel \cite{F-bbA} when no $b$--quarks
are detected in the final state.}.\s

Recently, the processes $pp \to gg+b\bar b \to \Phi \to \tau^+\tau^-$ have been
analyzed for the lHC \cite{Hpaper}: the production cross sections and the decay 
branching fractions have been updated (for the cross section part, see also
Ref.~\cite{LHCXS}) and the associated theoretical uncertainties, which turned
out to be quite large, have been discussed in detail. Relying on this analysis,
we will show in the present note that:\s

$i)$ The ATLAS and CMS exclusion limits in the $[M_A, \tb]$ parameter space,
which have been presented in the so--called $M_h^{\rm max}$ maximal mixing
benchmark scenario that maximizes the $h$ boson mass \cite{benchmark}, are in
fact almost model independent as  the supersymmetric particle spectrum enters
mainly through a radiative correction to the Higgs--$b\bar b$ Yukawa coupling
that essentially cancels out in the production cross section times decay
branching ratio. \s

$ii)$ If the ATLAS and CMS limits are applied to the minimal  predicted $pp\to
\Phi \to \tau^+ \tau^-$ rate when the theoretical uncertainties  are properly
taken into account, the excluded  range in the $[M_A, \tb]$ plane is slightly
smaller than indicated: only values $\tan\beta \gsim 30$--50 are excluded  in
the mass range 100 GeV $\lsim M_A \lsim 200$ GeV.  \s

$iii)$ These ATLAS and CMS exclusion limits will be significantly improved with
additional data and, for the luminosity of 1--3 fb$^{-1}$ expected at the end of
the $\sqrt s=7$ lHC run, they will be so strong that all other neutral and
charged MSSM Higgs search channels  will add only a modest improvement and will  
not be relevant  anymore for discovery.\s 

Finally, and most important, we point out that using the present analyses but
with a few inverse femtobarn accumulated data, the search  for the SM Higgs
particle in the process $gg\!\to\! H\! \to\! \tau^+ \tau^-$, which is not
currently considered by the  ATLAS and CMS collaborations,  might prove to be
very promising and a Higgs discovery would be possible in the otherwise rather
difficult mass range $115\;{\rm GeV}\! \lsim\! M_H\! \lsim\! 135$ GeV at the 
end of the early LHC run. 


\subsection*{2. $\mathbf{pp\! \to\!\Phi\! \to \!\tau^+ \tau^-}$ production rates
and model independence} 

The evaluation of the cross sections in the $gg \to \Phi$ and $b\bar b  \to
\Phi$ production processes at the lHC has been discussed in detail in 
Ref.~\cite{Hpaper} and we will only summarize here the main lines. Concentrating
on the pseudoscalar $A$ boson case at high $\tb$ values, $\sigma (gg \to A)$
which is known up to  next-to-leading order (NLO) only \cite{SDGZ} is
calculated  using the program {\tt HIGLU} \cite{Michael} with central values for
the renormalization and factorization scales\footnote{Our central scale is the
same as the one adopted in the SM Higgs case and is thus different from that
used in Ref.~\cite{LHCXS} where $\mu_0=M_H$ has been chosen; we thus obtain a
$gg\to A$ cross section that is $\approx 10\%$ larger.},
$\mu_R\!=\!\mu_F\!=\!\mu_0\!= \!\frac12 M_A$; only the very strongly enhanced 
loop contribution of the bottom quark is taken into account.  For the  $b\bar
b\to A$ process, known  up to next-to-next-to-leading order (NNLO)
\cite{bbH-NNLO}, we use the program {\tt bbh@nnlo}  \cite{Robert} with a central
scale $\mu_R\!=\mu_F\!= \!\mu_0\!=\! \frac14 M_A$. In both cases, the
$\overline{\rm MS}$ scheme for the renormalization  of the $b$--quark mass is
adopted\footnote{However, while the value $\overline{m}_b (\overline{m}_b)$ is 
used in the $gg$ process,  $\overline{m}_b(\mu_R)$ is adopted in the $b\bar b$
channel.}. The resulting partonic cross sections are then folded with the latest
MSTW sets of PDFs \cite{MSTW}, consistently at the respective perturbative 
orders. The cross sections are then multiplied by the $A\to \tau^+ \tau^-$ decay
branching fraction that we evaluate using the program {\tt HDECAY} \cite{HDECAY}
in which all (suppressed) channels except for $A\to b\bar b$ and $\tau^+\tau^-$
are ignored, leading to a value BR$(A\to \tau^+ \tau^-)=m_\tau^2/[3
\overline{m}_b^2(M_A)+m_\tau^2]\approx 10\%$.\s

In both production and decay processes, we assume the $b\bar bA$ coupling to be
SM--like, $\lambda_{Abb}= m_b /v$. To obtain the true cross  sections, one has
to rescale the obtained numbers by a factor  $\tan^2 \beta$. In addition, to
obtain the cross section for both the $A$ and $H(h)$ bosons, an  additional
factor of two has to be included. In most cases, this turns out to be a very
good approximation\footnote{This approximation is very  useful in practice as it
prevents the need of large grids to tackle numerically every MSSM scenario as
well as CPU time consuming  scans of the supersymmetric parameter space.} for
the  following reasons: \vspace*{1mm}

$i)$ As a consequence of chiral symmetry for $M_\Phi\! \gg\!\overline{m}_b$  
and because the Higgs masses and couplings are very similar,  the production and
decay amplitudes are the same for $A$ and $H(h)$. The only exception to our
simple rule is at masses $M_A\!\approx\! M_h^{\rm max}$. In this case, we are
not anymore in the decoupling or antidecoupling regimes, but in the so--called
intense coupling regime \cite{S-intense} in which the three neutral Higgs bosons
have comparable masses and similarly enhanced couplings to $b$--quarks.  As the
squares of the CP--even Higgs couplings  add to the square of the CP--odd Higgs
coupling, and since $M_H\! \approx\! M_h\!\approx M_A$, our results are
recovered provided that the cross section times branching ratios for the three
$h,H,A$ particles are added.\s

$ii)$ As the pseudoscalar $A$ boson does not couple to squarks of the same
flavor ($A \tilde q_i \tilde q_i$ couplings are forbidden by CP--invariance),
there is no superparticle contribution in the $gg \to A$ process at leading order and
higher order corrections are suppressed. In the CP--even Higgs case, there are
additional  superparticle contributions to $gg\to H(h)$ originating from (mainly stop and
sbottom) squark loops. However,  these contributions are damped by the squark
mass squared and are not similarly  enhanced by $ m_b \tb$ factors; they thus 
remain small so that they can be safely neglected in most cases.\s

$iii)$ The only relevant effect of the supersymmetric particles appears
through   the one--loop  vertex  correction  $\Delta_b$ \cite{Deltab} to the
$\Phi b\bar b$ coupling  which can be significant as it grows with $\tan\beta$. 
However, in the case of $p p\! \to\! \Phi\! \to\! \tau^+ \tau^-$, this
correction   almost entirely cancels out in the cross section times branching 
ratios and the remaining part is so small that it has no practical impact
whatever benchmark scenario is considered. Indeed,  the $\Delta_b$ correction
induces a shift
\beq
\sigma \times {\rm BR} \to \frac{ \sigma}{(1+\Delta_b)^2} \times \frac{\Gamma(\Phi \to
 \tau \tau)} { (1+\Delta_b)^{-2} \Gamma(\Phi\to b\bar b)+ \Gamma(\Phi \to \tau\tau)} 
 \approx  \sigma \times {\rm BR} \times (1- \frac15 \Delta_b)
 \eeq
assuming BR$(\Phi \to \tau^+ \tau^-)\approx 10\%$. Thus, unless the
$\Delta_b$ 
correction is extremely large, it will lead to only a few percent correction  at
most to the cross section times decay branching, which is negligible in view
of the much larger QCD uncertainties as will be discussed later.  \s

\begin{figure}[!h]
\begin{center}
\epsfig{file=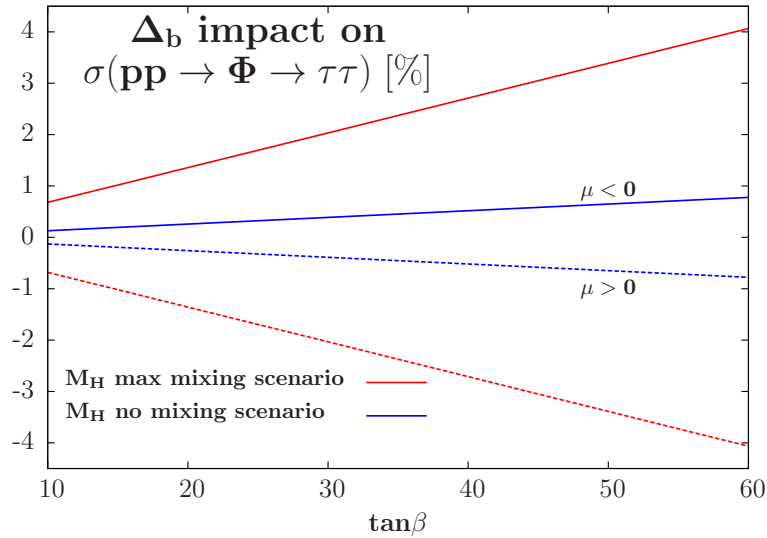,width=10cm} 
\end{center}
\vspace*{-6mm}
\caption[]{The impact (in \%) of the $\Delta_b$ supersymmetric radiative
correction on the cross section times branching ratio $\sigma[pp\to A+H(h)]
\times{\rm BR}[A /H(h) \to \tau^+\tau^-]$ as a function if $\tb$ in two of the 
benchmark scenarios of Ref.~\cite{benchmark} for both signs of $\mu$.}
\vspace*{-2mm}
\label{approximation}
\end{figure}

This feature is illustrated in Fig.~1 where, using the program FeynHiggs
\cite{feynhiggs} to evaluate the $\Delta_b$ correction, we display for a fixed 
value of $M_A$ and as a function of $\tan\beta$, the impact of the $\Delta_b$
correction on $\sigma (gg+b\bar b\! \to \Phi) \times  {\rm BR}(\Phi \to \tau
\tau)$. This is done in two benchmark scenarios for the CP--conserving  MSSM
proposed in Ref.~\cite{benchmark}: the maximal $M_h^{\rm max}$ mixing and  the
$M_h^{\rm min}$ no--mixing Higgs scenarios\footnote{The gluophobic  scenario is
now ruled out as it leads to light gluinos $m_{\tilde g} =500$ GeV) and squarks
($m_{\tilde q} \approx 350$ GeV) which have been  excluded by the recent ATLAS and
CMS analyses \cite{LHC-SUSY}. The small $\alpha_{\rm eff}$, which leads to
$\tilde m_g =500$ GeV and $\tilde m_q =800$ GeV is probably also excluded, in
particular if the various ATLAS and CMS analyses are combined. This latter 
scenario leads to a huge (and probably rather problematic) $\Delta_b$ value but
the effect on the cross section times branching ratio is again less than 10\%
for $\tb \lsim 30$.}, with the two possible signs of the higgsino parameter
$\mu$.\s

As can be seen, in both cases the quality of our approximation for  the
$pp\!\to\! \Phi\! \to\! \tau^+\tau^-$ cross section is always very good, the
difference with the exact result including the $\Delta_b$ correction being less
than 2\% for $\tb \lsim 30$ (and $\lsim 4\%$ for $\tb \lsim 60$),
which is negligible in view of the large QCD uncertainties that affect the cross
section.\s

Thus, our approximation is very good and the exclusion limit derived from the
ATLAS and CMS analyses depend very little on the supersymmetric model under
consideration. In fact, they also hold in a general  two--Higgs doublet model in
which two Higgs particles have the same mass and the same enhanced couplings to
isospin down--type  fermions. 

\subsection*{3. Impact of the theoretical uncertainties on the exclusion
limits} 

We turn now to the discussion of the ATLAS and CMS limits in the light of the
theoretical uncertainties  that affect the Higgs production cross section and
the decay branching ratios. These uncertainties have been discussed in detail in
Ref.~\cite{Hpaper} and can be  summarized as follows, starting with the   $gg\to
A$ and $b\bar b \to A$ production cross sections.  \s

-- The uncertainty from the missing higher orders in perturbation theory is
estimated by varying the renormalization and factorization scales in the domains
$ \mu_0/\kappa \le \mu_R,\mu_F  \le \kappa \mu_0$ around the central scales
$\mu_0$, with the additional restriction $1/\kappa \le \mu_R/\mu_F \le \kappa$
imposed.  While we choose $\kappa\!=\!2$ for the $gg\to A$ process,  the value
$\kappa\!=\!3$ is adopted for $b\bar b \to A$. \s

-- In $gg\to A$, an additional uncertainty is due to the choice of the scheme
for the renormalization of the $b$--quark mass which is estimated by taking the
difference between the results obtained in the on--shell mass and $\overline{\rm
MS}$ schemes and allowing for both signs. In $b\bar b \to A$, the inclusion of
this  effect is similar to increasing the  domain of scale variation from
$\kappa\!=\!2$ to $\kappa\!=\!3$.\s

-- The combined uncertainty from the PDFs and the  coupling $\alpha_s$ are
estimated within the MSTW scheme \cite{MSTW} by considering the PDF+$\Delta^{\rm
exp}\alpha_s$  uncertainty  at the 90\% CL to which we add in quadrature the
impact of a theoretical error on $\alpha_s$ as estimated by MSTW,  $\Delta^{\rm
th} \alpha_s\! \approx\! 0.002$ at NNLO.  We also add in quadrature a small 
uncertainty due to the $b$--quark pole mass value, $M_b=4.75 \pm 0.25$ GeV,  in the
$b$--quark and gluon densities.\s

-- Finally, there is the parametric uncertainty from the $b$--quark mass,
$\overline{m}_b (\overline{m}_b)=4.19^{+0.18}_{-0.06}$  GeV \cite{PDG} and, for
$b\bar b  \to A$, from $\alpha_s(M_Z^2)\!=\!0.1171\! \pm\! 0.0014$ at NNLO
\cite{MSTW}.\s

The uncertainties are combined as follows (see Refs.~\cite{Hpaper,Hpaper0} for
the argumentation): the PDF+$\Delta \alpha_s$+$\Delta m_b$ uncertainty are
evaluated on the minimal and maximal values of the cross sections with respect
to scale and scheme variation\footnote{ This procedure gives results that are
similar to those obtained with a linear addition of the scale+scheme and
PDF+$\Delta \alpha_s$+$\Delta m_b$ uncertainties as advocated  in, for instance,
Ref.~\cite{LHCXS}. Note that in this reference, from which the ATLAS and CMS
estimate of the uncertainty is borrowed, only the scale and  PDF+$\alpha_s^{\rm
exp}$ uncertainties are considered leading to a smaller total uncertainty than
the one we assume in our study.}; to this,  we add linearly the parametric
uncertainty on $\overline{m}_b$ which will drop anyway in the final result (see
below).\s 

\begin{figure}[!h] 
\begin{center} 
\mbox{
\epsfig{file=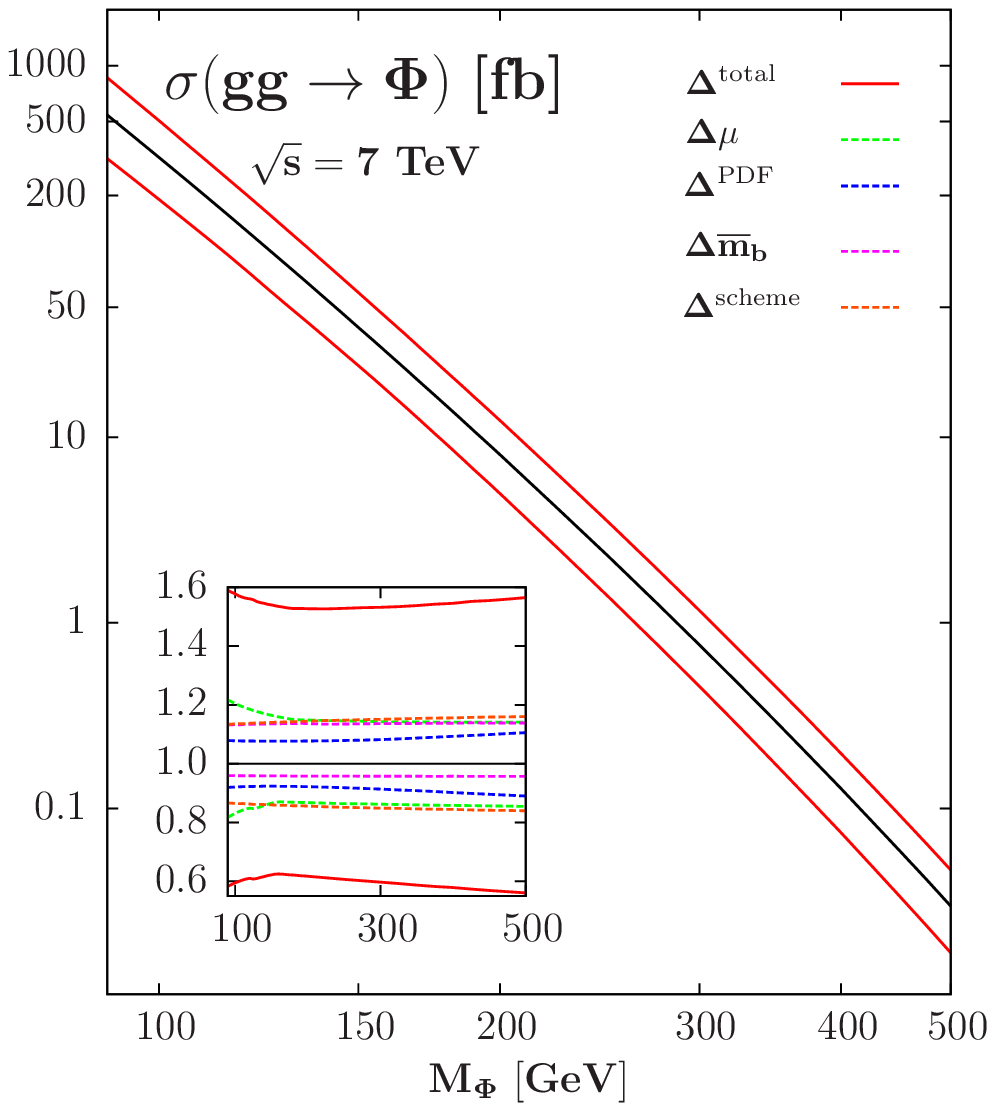,width=5.4cm}\hspace*{.2mm}
\epsfig{file=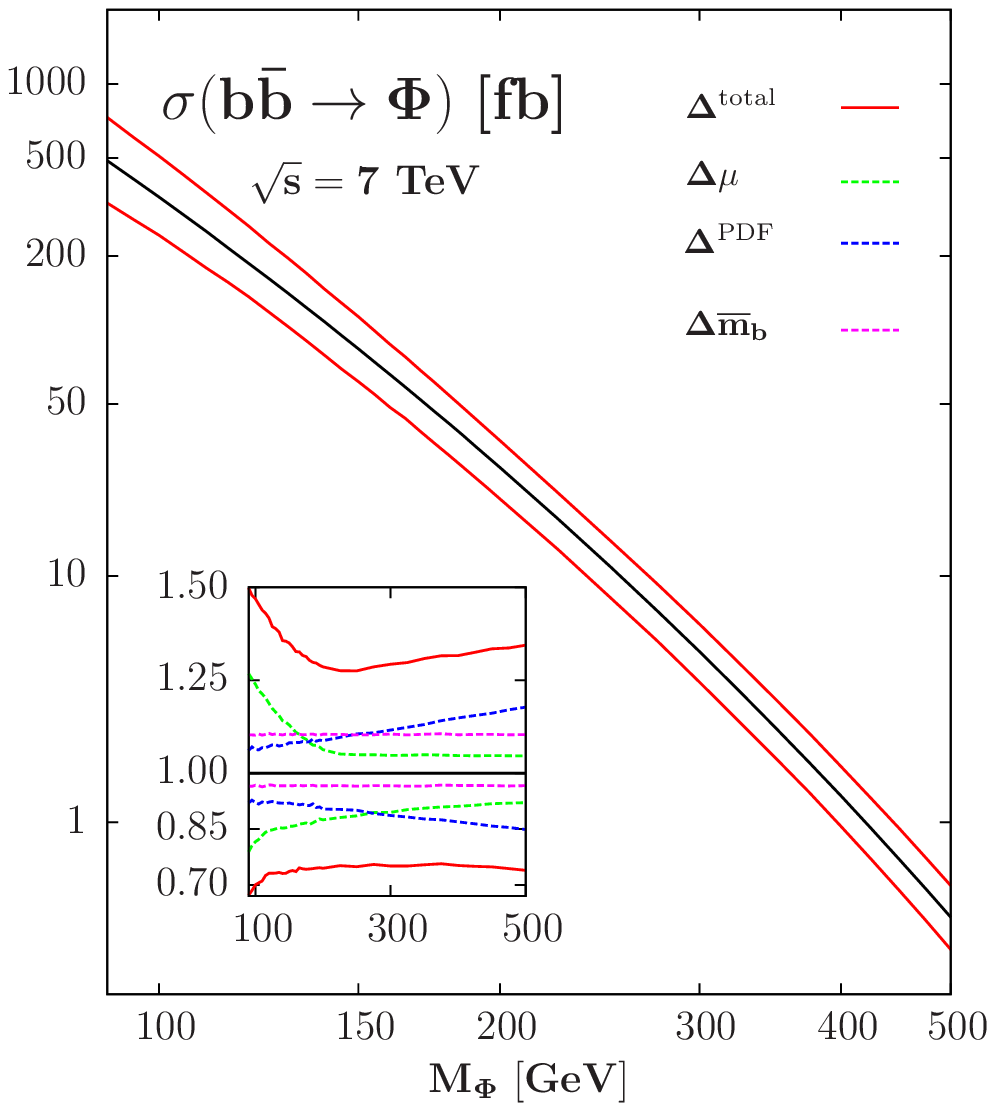,width=5.4cm}\hspace*{.2mm}  } 
\epsfig{file=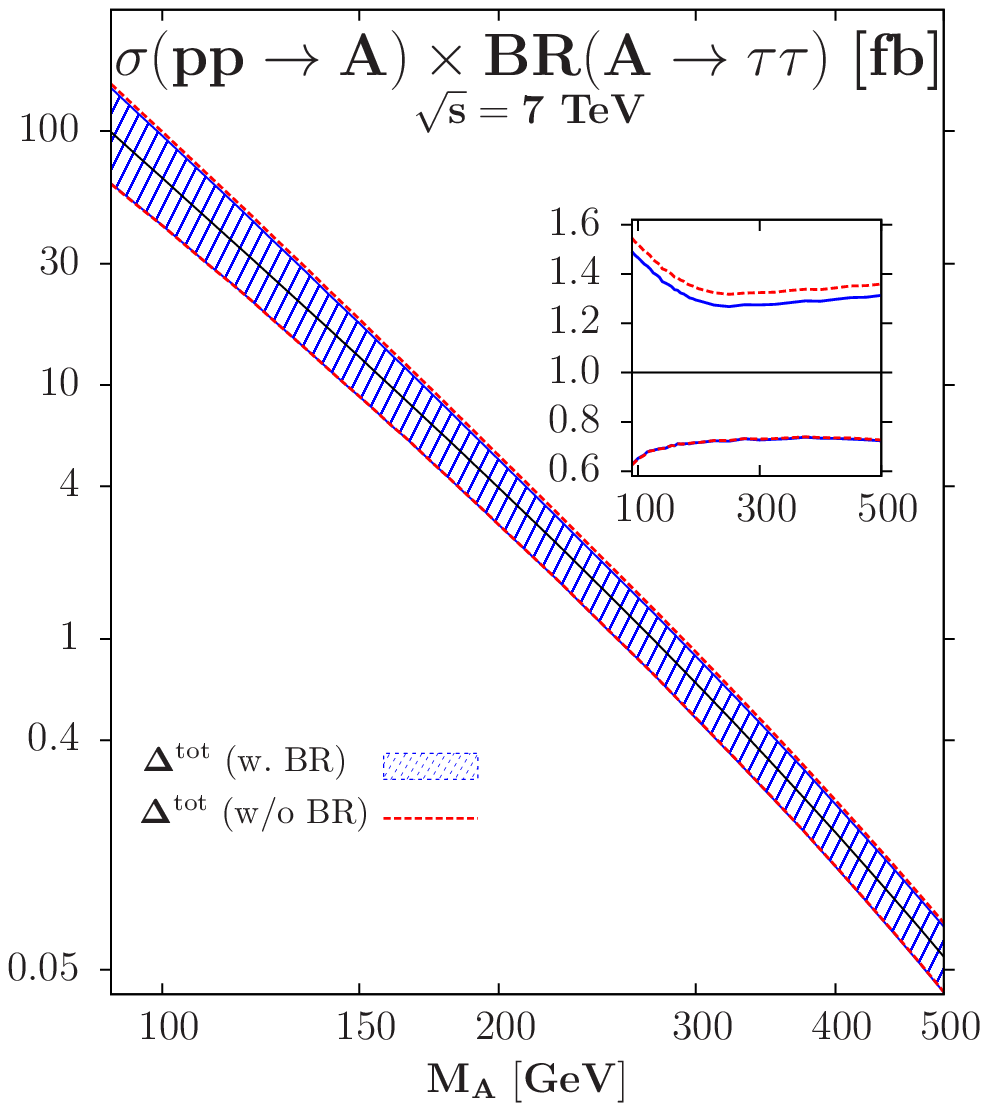,width=5.4cm} 
\end{center} 
\vspace*{-5mm}
\caption[]{The cross sections $\sigma^{\rm NLO}_{gg\!\to \!A}$ (left) and 
$\sigma^{\rm NNLO}_{b\bar b \!\to\!A}$ (center) at lHC energies as a function 
of $M_A$ when using the MSTW PDFs and unit $Ab\bar b$ couplings and the various
individual and total uncertainties. The combined $\sigma(p p\! \to\! A)\times 
{\rm BR}(A \!\to\! \tau^+ \tau^-)$ total theoretical uncertainties with and 
without the branching ratio is shown in the right panel. In the inserts, shown 
are the various theoretical uncertainties when the rates are normalized to  
the central values.}
\vspace*{-3mm}
\label{Errors}
\end{figure}

The results for the individual and total  uncertainties on the production cross
sections at the lHC are shown in Fig.~\ref{Errors} (left and central) for the
$gg\!\to\! A$ and $b\bar b  \!\to \!A$ processes as a function of $M_A$. One can
see that a total uncertainty of $\approx +60\%,-40\%$ for $\sigma(gg\to \Phi)$
and $\approx +50\%,-30\%$ for $\sigma(b\bar b \to \Phi)$ are obtained in the 
100--300 GeV Higgs mass range.\s  

Finally, the total uncertainty on the cross section times branching ratio 
$\sigma (pp \to \tau^+ \tau^-)$ is obtained by adding  the total  uncertainties
on the two production cross sections\footnote{We simply add all uncertainties
linearly, in contrast to Ref.~\cite{LHCXS} in which the PDF uncertainties in $gg$
and $b\bar b \to A$ are added in quadrature, with the total PDF+$\alpha_s$ 
uncertainty added linearly to the scale uncertainty.} and the uncertainties on
the branching fraction in Higgs decays into $\tau^+ \tau^-$ pairs. The latter is
simply affected by the parametric uncertainties on the input $b$--quark mass and
the value of $\alpha_s$; one finds an uncertainty of $\approx +4\%,-9\%$ on
BR$(\Phi\!  \to\! \tau^+\tau^-)$ at the 1$\sigma$ level over the entire 
relevant Higgs mass range, $M_A=100$--300 GeV. When included, the latter
uncertainties will cancel the parametric uncertainties  in the cross section
generating a slightly smaller total uncertainty in the  cross section times
branching ratio compared to the cross section  alone. This is exemplified in the
right hand--side of Fig.~\ref{Errors} where both  total uncertainties are
displayed. \s

To illustrate the impact of these theoretical uncertainties on the MSSM [$M_A,
\tan\beta$] parameter space that has been  probed at the lHC  in the  $p p \!
\to \! \Phi \! \to \! \tau^+ \tau^-$ channel, we consider the ``observed'' 
values of the cross section times branching ratio that have been given by the
CMS  collaboration\footnote{Unfortunately, the ATLAS collaboration has not given
this important information in its note \cite{ATLAS}. But as the ATLAS exclusion
limits are similar to those obtained by the CMS collaboration, the final results
once the theory uncertainties have been included should be the same.} for the
various values of $M_A$ \cite{CMS} and turn them into exclusion limits in this
plane by simply rescaling $\sigma(gg+b\bar b \to A \to \tau\tau)$ by a factor
$2\times \tan^2 \beta$. \s

\begin{figure}[!h]
\begin{center}
\vspace*{-2mm}
\epsfig{file=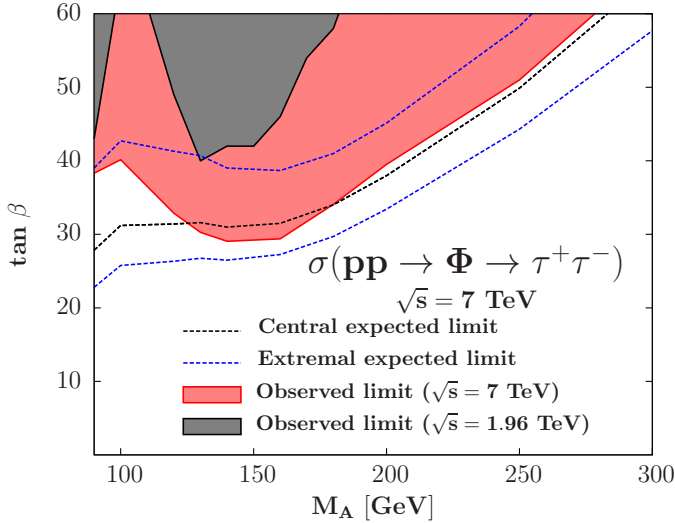,scale=0.8} 
\end{center}
\vspace*{-7mm}
\caption[]{Contours for the expected $\sigma(p p\! \to\! \Phi \!\to 
\!\tau^+ \tau^-)$ exclusion limits at the lHC in the [$M_A, \tb$] plane with 
the associated theory  uncertainties, confronted to the 95\% CL exclusion limits
given by the CMS \cite{CMS} and also CDF/D0 \cite{Tevatron} 
collaborations when our procedure is applied. } 
\vspace*{-5mm}
\label{Scan}
\end{figure}

This is shown in Fig.~\ref{Scan} where  the contour of the cross section times
branching ratio in this plane is displayed, together with the contours when the
uncertainties are included.  However, rather than applying the limits on the
central $\sigma \times$BR rate (as the CMS and also ATLAS collaborations do),
we  apply them on the minimal one  when the theory uncertainty is included.
Indeed, since the latter uncertainty is of theoretical nature, we will consider
that it has a flat prior and, hence, the minimal cross section times branching 
ratio  value is as respectable and likely as the central value.  In this case,
one observes that only values $\tb\gsim 28$ are  excluded for a Higgs  mass 
$M_\Phi \!\approx \!130$ GeV, compared to $\tb \gsim 23$ if the central
prediction is considered as in the CMS analysis. Hence, the inclusion of the
theory uncertainties should lead to a slight reduction of the excluded  [$M_A,
\tan\beta$] parameter space.

\subsection*{4. Implications in the MSSM for higher luminosities}

The exclusion limits on the $[\tb, M_A$] MSSM parameter space obtained by the
ATLAS and CMS collaborations with  only 36 pb$^{-1}$ data are extremely strong
as, for instance, values $\tb \gsim 30$ are excluded in the low mass range for
the pseudoscalar Higgs boson, $M_A=90$--200 GeV. This has several consequences
that we briefly summarize below. \s

\begin{figure}[!h]
\begin{center}
\vspace*{-2mm}
\epsfig{file=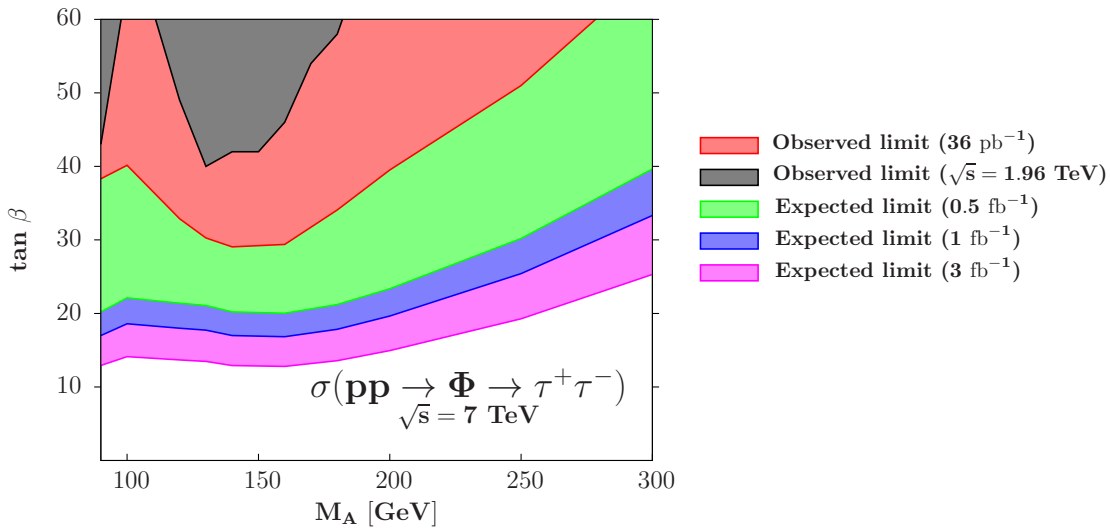,scale=0.8} 
\end{center}
\vspace*{-7mm}
\caption[]{Contours for the ``expected" $\sigma(p p\! \to\! \Phi \!\to 
\!\tau^+ \tau^-)$ 95\% CL exclusion limits at the lHC with $\sqrt s=7$ TeV in 
the [$M_A, \tb$] plane for various integrated luminosities. The present limits
from CMS and the Tevatron are also displayed. } 
\vspace*{-5mm}
\label{projection}
\end{figure}

First of all, if the luminosity is increased to the fb$^{-1}$ level as is
expected to be the case already at the end of this year, the values of $\tb$
which can be probed will be significantly lower. Assuming that there will be no
improvement in the analysis (which might be a little pessimistic as discussed
later) and that the CMS sensitivity will simply scale as the square root of the
integrated luminosity, the region of the $[\tb,M_A]$ parameter space which can
be excluded in the case where no signal is observed is displayed in 
Fig.~\ref{projection} for several values of the accumulated luminosity. With 3
fb$^{-1}$ data per experiment (or with 1.5 fb$^{-1}$ when the ATLAS and CMS
results are combined), values $\tb \gsim 12$ could be excluded in the entire mass
range $M_A \lsim 200$ GeV; the exclusion reduces to $\tb \gsim 20$ for the mass
range $M_A \lsim 300$ GeV. \s

These limits could be improved by considering four additional production
channels.\s

$i)$ The process  $gb \to \Phi b \to b b\bar b$ where the all final bottom
quarks are detected: the production cross section $\sigma(b g \to b\Phi+g\bar
b\to \Phi \bar b)$ is one order of magnitude lower than that of the inclusive
$gg+b\bar b\to \Phi$ process (the $bg\to  b \Phi$ process is part of the NLO
corrections to $b\bar b \to \Phi$ \cite{bbH-NLO}) but this is compensated by the
larger fraction BR$(\Phi \to b\bar b) \approx 90\%$ compared to BR$(\Phi \to
\tau^+ \tau^-) \approx 10\%$; the QCD background are much larger
though\footnote{We estimate the theoretical uncertainties on the rate $\sigma(bg
\to Ab \to b b\bar b)$  to be similar to that of the combined $gg+b \bar b \to
\Phi$ process, that is $\pm 30$--40\%. Note, however,  that here the 
parametric and supersymmetric  corrections do not cancel in the cross section
times branching ratio and have to be taken into account.}.\s

$ii)$ The process  $pp \to \Phi \to \mu^+\mu^-$ for which the rate is simply
given  by $\sigma(pp\!\to\! \Phi \to \tau\tau )$ rescaled by BR$(\Phi \to \mu
\mu)/$BR$(\Phi \to \tau \tau)= m_\mu^2/m_\tau^2 \approx 4\times 10^{-3}$; the
smallness of the rate\footnote{Note that here, the  parametric and 
supersymmetric corrections cancel out exactly in the cross section times
branching ratio and we are left only with the scale/scheme and PDF+$\alpha_s$
uncertainties on the cross section that are exactly the same as those  discussed
previously.} is partly compensated by the much cleaner $\mu \mu$ final state and
the better resolution on the $\mu\mu$ invariant mass. The efficiency of the  $pp
\to \Phi\to \tau \tau$ signal is estimated by the CMS collaboration (see Table 1
of Ref.~\cite{CMS}) to be   4.5\% when all $\tau$ decay channels are considered.
This is only a factor 10 larger than the ratio BR$(\Phi \to \mu \mu)/$BR$(\Phi
\to \tau \tau)$ (which  is approximately equal to the efficiency in the $\tau
\to e\mu$  channel). Thus the $\Phi \to \mu \mu$ decay channel might be 
useful. In particular,  the small resolution that can achieved could allow to
separate the three peaks of the almost degenerate $h,H$ and $A$ states in the
intense coupling regime; see Ref.~\cite{S-intense}.\s

$iii)$  The process $pp \to tbH^- \to tb \tau \nu$ which leads to a cross
section that is also proportional to $\tan^2\beta$ (and which might also be
useful for very low $\tb$  values) but that is two orders of magnitude smaller 
than $\sigma(pp \to \Phi)$ for $M_A \approx 100$--300  GeV. \s

$iv)$ Charged Higgs production from top quark decays, $pp\! \to\! t\bar t$ with 
$t\to H^+b \to \tau^+ \nu b$, which has also been recently analyzed by the CMS
collaboration \cite{CMS-H+}. With 36 pb$^{-1}$ data, values of the branching
ratio  BR$(t\to H^+b) \gsim 25\%$ are excluded which means that  only $\tb$
values larger than 60 are probed for the time being\footnote{We note that in
this case the theoretical  uncertainties have not been estimated in
Ref.~\cite{LHCXS} (contrary to the channel  $pp\! \to\! tbH^-\! \to\! tb \tau
\nu$ where an uncertainty of $\pm 30\%$  has been found). We have
evaluated them with {\tt HATHOR} \cite{Hathor}
and, in the production channel  $\sigma(p  p\! \to\! t\bar t)$ and for
$m_t\!=\! 173.3\! \pm  \! 1.1$GeV, we  find  $\sigma(p  p \to t\bar t)\!=
\!163~^{+2.5\%}_{-5.6\%}~{(\rm
factor~2~from~central~m_t~scale)}~^{+10.4\%}_{-10.1\%}~ ({\rm 
PDF}\!+\!\Delta^{\rm exp+th}\alpha_s@90\%{\rm CL})~{\pm 3.3\%}~ ( \Delta m_t)$
pb, which leads  using the procedure of Ref.~\cite{Hpaper0}  to a total
uncertainty of  $\Delta \sigma/\sigma= ^{+16\%}_{-19\%}$, i.e three times larger
than the one assumed in the CMS analysis. To that, one should add  the
uncertainty on the branching ratio BR$(t\to H^+b)$  for which the parametric
one  (from the input $\overline{m}_b$ and $\alpha_s$ values) is about $+10\%,
-4\%$ \cite{Hpaper}.}.  \s

 Nevertheless, in the four cases, the small rates  will allow only for a modest
improvement over the $pp \to \Phi \to \tau \tau$ signal or exclusion limits. In
fact, according to the (presumably by now outdated) projections of the ATLAS and
CMS collaborations \cite{ATLAS+CMS-TDR}  at the full LHC with $\sqrt s=14$ TeV
and 30 fb$^{-1}$ data, these processes are observable only for  not too large
values of $M_A$  and relatively high values of $\tan\beta$ ($\tb \gsim 20$) most
of which are already  excluded and the remaining part will be excluded if no
signal is observed  at the end of the present year. However, as is the case for
the $pp\to \tau \tau$ channel, some (hopefully significant) improvement over
these projections might be achieved.   

\subsection*{5. Implications for the Standard Model Higgs boson}

A very important consequence of the ATLAS and CMS  $pp\to \tau \tau$ inclusive 
analyses  is that they open the possibility of using this channel in the case of
the  Standard Model Higgs boson $H$. Indeed, in this case, the main production
process is by far the $gg\to H$ channel which dominantly proceeds via a top
quark loop with a small contribution of the bottom quark loop. The cross section
for this process has been discussed in detail in Refs.~\cite{Hpaper,LHCXS} and,
in the mass range $M_H\!=\!115$--140 GeV, it is at the level of 10 to 20 pb. The
branching ratio for the decay  $H\! \to\! \tau^+\tau^-$ ranges from 8\% at
$M_H\!=\!115$ GeV to 4\% at $M_H\!=\!140$ GeV. The cross section times
branching ratio $\sigma(gg\!\to\! H\! \to \! \tau^+\tau^-)$ is thus rather
substantial at low Higgs masses\footnote{The cross section in the SM is
comparable to that of $A+H(h)$ production in the MSSM with values $\tan\beta
\approx 4$ (and not $\tb=1$!), a consequence of the dominance of the top--quark
loop (in the SM)  compared to bottom--quark loop (as is in general the case in
the MSSM) in the $gg$ fusion process.}.  \s

\begin{figure}[!h]
\begin{center}
\vspace*{-2mm}
\epsfig{file=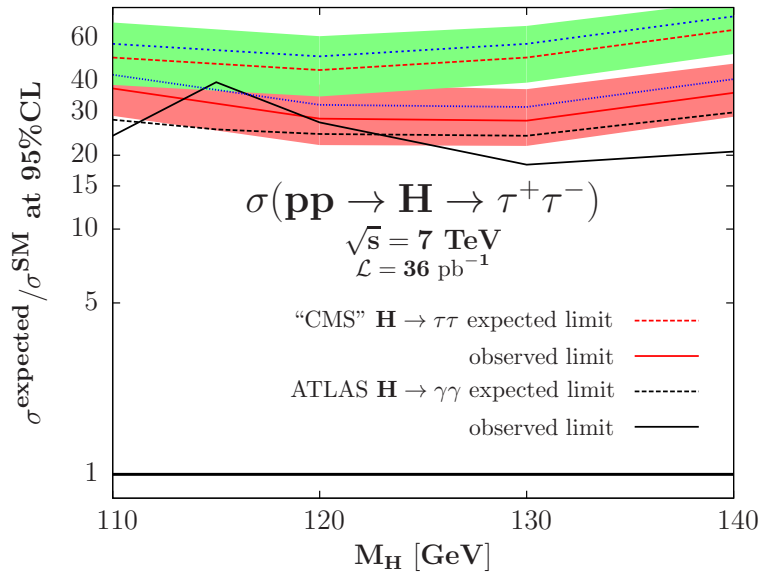,scale=0.89} 
\end{center}
\vspace*{-6mm}
\caption[]{The median expected and observed cross sections at 95\% CL for the 
production of the SM Higgs boson in the channels $gg\! \to\! H\to \!\tau \tau$ 
(blue lines) and $pp\! \to\! H \to \!\tau^+\tau^-+X$ (red lines captioned) from an 
extrapolation of the CMS analysis with 36 pb$^{-1}$ data \cite{CMS} normalised
to the SM cross section. It is compared to the case of $H\!\to \!\gamma \gamma$
as recently analyzed by the  ATLAS collaboration  \cite{ATLAS-gamma} with 37.6
pb$^{-1}$ data.}
\vspace*{-2mm}
\label{SMHiggs}
\end{figure}

Using the numerical values of the SM Higgs cross sections and the decay
branching ratio of Ref.~\cite{Hpaper} as well as  the ``median expected" and
``observed"  95\%CL limits obtained in the CMS analysis at $\sqrt s=7$ TeV with
36 pb$^{-1}$ data, we display in Fig.~\ref{SMHiggs} the ratio  of the observed
and expected cross sections at the 95\%CL normalised to the  SM cross section
$\sigma(gg\to H\to \tau^+\tau^-)$ as a function of the Higgs mass. One can see
that with the small amount of data available today,  we are a factor $\approx
50$ to 60 above the expected rate in the SM in the mass range $M_H=110$--140
GeV.   However, when we compare this rate to the expected and observed rates
(taken from a recent ATLAS analysis with 37.6 pb$^{-1}$ data \cite{ATLAS-gamma})
for the most important and promising channel for the SM Higgs boson in this mass
range, $H\to \gamma \gamma$, one sees that the situation is not that bad. 
Indeed, the $gg\to H\to \tau\tau$ inclusive channel, which has not been
considered neither by the ATLAS nor the CMS collaborations, is in fact rather
powerful and competes rather well with the long celebrated  $H\to \gamma \gamma$
detection channel, as  it has a sensitivity that is only a factor of two smaller
than the latter channel. \s

Thus, the $gg\!\to\! H\!\to\! \tau^+\tau^-$ channel could be also used to search
for the SM Higgs  in the very difficult mass range $M_H\! \approx\! 115$--130
GeV where only the $H\to \gamma \gamma$ channel was considered to be effective.
The two channels could be combined to reach a better sensitivity. In addition,
while little improvement should be expected in $H\to \gamma \gamma$
(for which the analyses have been tuned and optimised for more than twenty
years now), a better sensitivity could be achieved in the  $H\to \tau^+\tau^-$
channel. Indeed, on the one hand,  a possible improvement might come from the
experimental side: novel and better mass reconstruction techniques of the 
$\tau\tau$ resonance can be used\footnote{In Ref.~\cite{Mtau}, a new mass
technique for reconstructing resonances decaying into $\tau$ lepton pairs has
been proposed and  it is claimed that it allows for a major improvement in the
search for the Higgs$\to\! \tau\tau$ signal.}, splitting of  the analysis into 
jet multiplicities as done for instance for the channel $H\! \to\! WW \to \ell
\ell \nu \nu$, inclusion of additional topologies such as same sign $\ell=
e,\mu$ final states,  etc... On the other hand, one can render the $H\to \tau
\tau$ channel more effective by simply including the contribution of the other
Higgs production mechanisms  such as vector boson fusion\footnote{In fact, there
are separate analyses  of the vector boson fusion process with $H\to \tau \tau$
(first proposed in Ref.~\cite{VBF-tautau}) made by the ATLAS and CMS
collaborations \cite{ATLAS+CMS-TDR}. Because of the two additional forward jets,
the sensitivity of this channel (the only one involving $\tau$ leptons that has
been considered in the SM so far) is much larger than its contribution ($\approx
10\%$) to the total $pp\to H\to \tau \tau+X$ inclusive rate indicates.}   and 
associated production with a $W$ and $Z$ bosons, which will increase the cross
section for the inclusive $pp\to \tau \tau+X$ production  mechanism by 15 to
20\%. This is exemplified in Fig.~~\ref{SMHiggs} where the sensitivity is shown
when the additional contributions of these processes are (naively, i.e. without
making  use of the specific cuts for vector boson fusion which will
significantly increase the sensitivity) included in the cross section for the 
inclusive $pp\to H\to \tau^+ \tau^-+X$ signal.\s

Hence, the $pp\to {\rm Higgs} \to \tau^+\tau^-$ inclusive channel turns out to
be a very interesting and potentially very competitive Higgs detection channel
also in the Standard Model and it should be considered with a higher priority by
the ATLAS and CMS collaborations. 

\subsection*{6. Conclusions} 

We have discussed the implications of the recent analyses performed by the ATLAS
and CMS collaborations in the $pp \to \tau^+ \tau^-$ search for the MSSM neutral
Higgs bosons with 36 pb$^{-1}$ data. The results lead to very strong constraints
on the $[\tan\beta, M_A]$ MSSM parameter space. We have shown that these
constraints are essentially model independent (for the values of $M_A$ and $\tb$
that are being probed  so far) and slightly less effective when the theoretical
uncertainties in the predictions for the Higgs cross sections and branching
ratios are properly included.   If a Higgs signal is still  absent with a few
inverse femtobarn data, these limits  can be significantly improved  and values
$\tb \lsim 10$ can be excluded for not too heavy neutral Higgs  bosons. At this
stage, many channels such as  $pp\! \to\! \Phi\! \to \mu^+ \mu^-, pp \!\to
\!btH^- \to\! bt \tau\nu$, $gb \!\to \!\Phi b\!\to \!3b$ and $t \to H^+b$ will
not be viable  anymore even at the design energy and luminosity of the LHC.\s

The most important remark that we make in this note is that the  inclusive
$pp\to {\rm Higgs} \to \tau \tau$ process is also a very promising channel in
the search for the Standard Model Higgs boson. Indeed, this channel has a
sensitivity that is  only a factor of two smaller than the expected sensitivity
of the main  $pp \to H \to \gamma \gamma$ channel. While little improvement is
expected in the later channel, there are ways to significantly enhance the
sensitivity of the $pp\to H\to \tau\tau$ signal and render it a very powerful
discovery  channel for the SM Higgs boson in the difficult $M_H=115$--130 GeV
mass range. We thus urge our experimental colleagues from the ATLAS and CMS
collaborations to very seriously consider this additional and promising
possibility.  \bigskip

{\bf Acknowledgments:} Discussions with Ketevi Assamagan, Sasha Nikitenko and
Markus Schumacher on the ATLAS and CMS analyses  are gratefully acknowledged.
This work is supported by the  ERC Grant ``Mass Hierarchy and Particle
Physics at the TeV Scale".

\end{document}